\documentclass[12pt,titlepage]{article}

\font\grande=cmr9.5 scaled \magstep4
\font\medio=cmr9.5 scaled \magstep2
\outer\def\beginsection#1\par{\medbreak\bigskip
      \message{#1}\leftline{\bf#1}\nobreak\medskip
\vskip-\parskip
      \noindent}
\usepackage{graphicx} 
\begin{document}
\bibliographystyle {unsrt}

\titlepage

\begin{flushright}
CERN-PH-TH/2015-068
\end{flushright}

\vspace{1cm}
\begin{center}
{\grande Non-linear curvature inhomogeneities}\\
\vspace{5mm}
{\grande and backreaction for relativistic viscous fluids}\\
\vspace{1cm}
 Massimo Giovannini 
 \footnote{Electronic address: massimo.giovannini@cern.ch} \\
\vspace{1cm}
{{\sl Department of Physics, 
Theory Division, CERN, 1211 Geneva 23, Switzerland }}\\
\vspace{0.5cm}
{{\sl INFN, Section of Milan-Bicocca, 20126 Milan, Italy}}
\vspace*{1cm}
\end{center}

\vskip 0.3cm
\centerline{\medio  Abstract}
\vskip 0.1cm
The non-perturbative curvature inhomogeneities induced by relativistic viscous fluids are not conserved in the large-scale limit. However when the bulk viscosity is a function of the total energy density of the plasma (or of the trace of the extrinsic curvature) the relevant evolution equations develop a further symmetry preventing the non-linear growth of curvature perturbations. In this situation 
the fully inhomogeneous evolution can be solved to leading order in the gradient expansion. Over large-scales both the acceleration and the curvature inhomogeneities are determined by the bulk viscosity coefficients. Conversely the shear viscosity does not affect the evolution of the curvature and does not produce any acceleration. The curvature modes analyzed here do not depend on the choice of time 
hypersurfaces and are invariant for infinitesimal coordinate transformations in the perturbative regime. 
\noindent

\vspace{5mm}
\vfill
\newpage
\renewcommand{\theequation}{1.\arabic{equation}}
\setcounter{equation}{0}
\section{Introduction}
\label{sec1}
The temperature and the polarization anisotropies of the cosmic microwave background are adequately described in the framework of the linear theory (see, e.g. the WMAP or WMAP9 data essentially consistent with the Planck explorer data \cite{data1,data2,data3}). However throughout the whole history of the plasma a non-linear growth of the curvature inhomogeneities cannot be excluded. The initial conditions of the temperature and polarization anisotropies of the cosmic microwave background are customarily set after neutrino decoupling \cite{data1,data2,peebles} but the evolution of curvature perturbations starts at a much earlier epoch namely during inflation and possibly even before. It is then particularly interesting, from the theoretical viewpoint, to define and discuss
plausible non-perturbative generalisations of the curvature inhomogeneities in different physical situations going beyond the conventional 
cases of a single scalar field and of a perfect barotropic fluid. The theme of this paper concerns the non-perturbative generalization of the curvature inhomogeneities,  their evolution and their physical relevance when the energy-momentum tensor is dominated by relativistic viscous effects at large-scales.

Prior to the formulation of the inflationary paradigm the non-linear evolution of curvature perturbations has been analyzed in various frameworks. The approach followed here can be traced back to the expansion in spatial gradients of the geometry \cite{bel1,bel2} (see also \cite{LL}). The gradient expansion has been used either in the proximity of the big-bang singularity or away from it.  Close to the singularity the geometry may be highly anisotropic but it turns out to be rather homogeneous \cite{bel1,bel2}. As soon as an inflationary event horizon is formed \cite{star,salope,tomita}  any finite portion of the event horizon gradually loses the memory of an initially imposed anisotropy or inhomogeneity so that the metric attains the observed regularity regardless of the initial boundary conditions as hypothesized  in the past \cite{hoyle,zel1,misner}.  

One of the central themes of the inflationary paradigm \cite{inf1,inf2} is to wash out primeval anisotropies in the expansion right after the formation of the inflationary event horizon (see, however, Ref. \cite{barrow}). Probably the first non-linear generalization of inflationary curvature perturbations has been proposed in \cite{salope} after the pioneering analyses on the gauge-invariant treatment of linearised cosmological perturbations \cite{pio}.  These have been subsequently scrutinized and rediscovered by different authors \cite{shell}. The non-perturbative generalizations 
discussed here may also have some impact on the neighbouring problems such as the higher-order approaches to cosmological perturbations (see \cite{NH} and references therein). 

It is  appropriate to gauge the effects of the viscosity coefficients that can play a relevant role both in the early and in the late Universe.
Since this investigation addresses the evolution of the non-linear curvature perturbation in the 
relativistic theory of viscous fluids (see, for instance, \cite{visc1} and references therein) the assumption of the he strict reversibility of the system will be dropped. The shear viscosity suppresses exponentially the traceless part of the extrinsic 
curvature. The bulk viscosity enters the definition of the curvature perturbations 
and may cause their non-conservations for typical length-scales larger 
than the Hubble radius. While the bulk viscosity does affect directly the non-linear deceleration
parameter (possibly causing accelerated expansion) the opposite is true for the shear viscosity.

From the technical viewpoint the variables introduced in the present analysis do not depend on the choice of time hypersurfaces and they are exactly invariant for infinitesimal coordinate transformations in the perturbative regime. It will be argued that the gravitating viscous fluids lead, under certain conditions, to a further symmetry preventing the non-linear growth of curvature inhomogeneities. The same set of conditions will also be shown to be compatible with a quasi-de Sitter stage of expansion.
In the reversible limit our variables coincide with the ones conventionally defined in the case of a single scalar field or for 
a perfect barotropic fluid. The results pursued here can also be used to deduce the corrections to the linear theory without going through second or even higher-order calculations \cite{NH}.

Finally, in the single inflaton case the large-scale cosmological perturbations 
are sometimes treated within the so-called $\delta {\mathcal N}$ formalism \cite{salope,shell} where ${\mathcal N}$ denotes the inhomogeneous expansion rate integrated in time generalizing to the non-perturbative level the total number of inflationary efolds.  The presence of non-adiabatic fluctuations of the pressure make the formalism less appealing but we will show that, in some specific cases, 
the $\delta {\mathcal N}$ formalism can also be applied in the case of relativistic viscous fluid. 

The layout of the investigation is the following. In section \ref{sec2} we shall discuss the basic aspects 
of the geometry and of its coordinate transformations. The interplay between relativistic viscous fluids and general relativistic gradient expansion is addressed in section \ref{sec3}.  Section \ref{sec4} is devoted to the non-perturbative evolution of large-scale curvature inhomogeneities in the viscous case.  The perturbative limit of our considerations is discussed in section \ref{sec5}. Section \ref{sec6} contains the concluding remarks. 

\renewcommand{\theequation}{2.\arabic{equation}}
\setcounter{equation}{0}
\section{Nonlinear gauge transformations}
\label{sec2}
In the Arnowitt-Deser-Misner formalism \cite{ADM1} (ADM in what follows) the line element is expressed in terms of the 
conventional $(3+1)$-dimensional decomposition:
\begin{equation}
ds^2 = g_{\mu\nu}(\tau,\vec{x})\, dx^{\mu} \, dx^{\nu} =N^2 d\tau^2 - \gamma_{ij} (d x^{i} + N^{i} d\tau)   ( d x^{j} + N^{j} d\tau),
\label{adm1}
\end{equation}
where $N=N(\tau,\vec{x})$ denotes the lapse function, $N^{i}=N^{i}(\tau,\vec{x})$ is the shift vector and $\gamma_{ij}=\gamma_{ij}(\tau, \vec{x})$ is the spatial 
three metric\footnote{Note that the spatial indices are lowered and raised using $\gamma_{ij}(\tau,\vec{x})$. The Greek indices will take the values 
$(\mu,\, \nu )= 0,\, 1\, 2\, 3$.}.  In the ADM variables of Eq. (\ref{adm1}) the extrinsic curvature of the spatial slices 
(i.e. $K_{ij}(\tau,\vec{x})$) and the components of the Ricci tensor of the spatial slices (i.e. $r_{ij}(\tau,\vec{x})$) become:
\begin{eqnarray}
K_{ij}(\tau,\vec{x}) &=& \frac{1}{2 N} \biggl[- \partial_{\tau}\gamma_{ij} + \nabla_{i}N_{j} + \nabla_{j} N_{i} 
\biggr],
\label{ADM1a}\\
r_{ij}(\tau,\vec{x}) &=& \partial_{m} \, ^{(3)}\Gamma^{m}_{ij} -\partial_{j} ^{(3)}\Gamma_{i m}^{m} + ^{(3)}\Gamma_{i j}^{m} 
\,^{(3)}\Gamma_{m n}^{n} - ^{(3)}\Gamma_{j n}^{m} \,^{(3)}\Gamma_{i m}^{n},
\label{ADM1b}
\end{eqnarray}
where, for short,  $^{(3)}\nabla_{i}= \nabla_{i}$ is the covariant derivative defined\footnote{We warn the reader that this identification will be followed 
throughout the paper. According to some, this notation may lead to potential ambiguities but we hope that, with this note, confusions 
will be avoided. In this paper $\nabla_{i}$ will denote the covariant derivative on the spatial slices and not the spatial component of a (four-dimensional) covariant derivative.}
 with respect to the metric $\gamma_{ij}$; $\partial_{\tau}$ denotes a derivation with respect to the time coordinate 
$\tau$ and $^{(3)}\Gamma_{i j}^{m}$ are the Christoffel symbols computed from $\gamma_{ij}$. Note that 
$\Gamma_{ij}^{m} = ^{(3)}\Gamma_{i j}^{m}$ but only in the case $N_{i}=0$. From Eq. (\ref{adm1}) the unit time-like vector normal to the $x^{0} = \tau = {\mathrm constant}$ hypersurface is $n_{\mu} = (N, 0)$ and $n^{\mu} = (1/N, \, - N^{i}/N)$. The choices $N=1$ and $N_{i}=0$ corresponds to the
geodesic slicing leading to the Gaussian normal coordinates, a popular gauge in numerical relativity \cite{shap}. The condition  $N_{i}=0$ implies that coordinate observers coincide with normal observers: the normal vector $n^{\mu}$ has vanishing spatial component.
Without positing a specific gauge choice, Eqs. (\ref{ADM1a}) and (\ref{ADM1b}) lead to the following expressions for the components of the Ricci tensor:
\begin{eqnarray}
R_{0}^{0} &=&  
\frac{\partial_{\tau} K}{N}- \mathrm{Tr}K^2 + \frac{\nabla^2 N}{N} - \frac{N^{m}}{N} \nabla_{m} K + \frac{N^{q}}{N} \biggl(\nabla_{q} K - \nabla_{k} K^{k}_{q} \biggr),
\label{ADM7}\\
R_{i}^{0} &=& \frac{1}{N} \biggl(\nabla_{i} K - \nabla_{k} K^{k}_{i} \biggr),
\label{ADM8}\\
R_{i}^{j} &=& \frac{1}{N} \partial_{\tau} K_{i}^{j} - K K_{i}^{j} - r_{i}^{j} + \frac{1}{N} \nabla_{i} \nabla^{j} N - \frac {N^{m}}{N} \nabla_{m} K_{i}^{j}
\nonumber\\
&+& \frac{1}{N} \nabla_{m} N^{j} K^{m}_{i}-  \frac{1}{N} \nabla_{i} N^{m} K_{m}^{j} - \frac{N^{j}}{N} \biggl(\nabla_{i} K - \nabla_{k} K^{k}_{i} \biggr),
\label{ADM9}
\end{eqnarray}
where the notations $K = \gamma^{ij} K_{ij}$ and $\mathrm{Tr}K^2 = K_{i}^{j} K_{j}^{i}$ have been adopted.

Notice that in the linear theory, for infinitesimal coordinate transformations of the type:
\begin{equation}
x^{\mu} \to \tilde{x}^{\mu} = x^{\mu} + \epsilon^{\mu}, \qquad \epsilon^{\mu} = (\epsilon^{0} , \, \epsilon^{i}),
\label{INF}
\end{equation}
the metric fluctuations change as the Lie derivative in the direction $\epsilon_{\mu}$ \cite{pio} (see also section \ref{sec5}). 
The perturbative expansion underlying Eq. (\ref{INF}) assumes the separation of the geometry into a background value 
supplemented by a perturbation. The choice of the temporal gauge defines the spatial hypersurface of fixed coordinate time while the choice 
of the spatial gauge determines the worldlines of fixed spatial coordinates.  The coordinate system is completely 
specified when both $\epsilon^{0}$ and $\epsilon^{i}$ are assigned. It is however possible to define various 
sets of gauge-invariant variables that do not change under Eq. (\ref{INF}). In the linearized treatment, the gauge parameters are of the same order of the metric perturbations: as soon as, in some gauge,  the perturbation variables grow non-linear and affect the background geometry the linearised approximation is no longer tenable. 

The approach pursued here does not assume the validity of the perturbative expansion insofar as the geometry is not split into a background value supplemented by a perturbation with small amplitude: this treatment is arguably the most suitable for the unambiguous analysis of some backreaction problems. When the metric is not linearized around a specific background, the coordinate transformations will not necessarily be infinitesimal and 
shall be parametrized as $x^{\mu} \to x^{\mu} = Y^{\mu}(x)$ or, in more explicit terms, as:
\begin{equation}
\tau \to T= T(\tau, \vec{x}), \qquad x^{i} \to Y^{i} = Y^{i}(\tau, \vec{x}).
\label{adm2}
\end{equation}
Under the transformation of Eq. (\ref{adm2}) the metric components of Eq. (\ref{adm1}) will change as
\begin{equation}
g_{\alpha\beta}(\tau, \vec{x}) = G_{\mu\nu}(T,\vec{Y}) \, \biggl(\frac{\partial X^{\mu}}{\partial x^{\alpha}}\biggr) \,  \biggl(\frac{\partial X^{\nu}}{\partial x^{\beta}}\biggr).
\label{admin3}
\end{equation}
The explicit form of Eq. (\ref{admin3}) can be written, schematically, as\footnote{The partial derivatives with respect to $\tau$ and $x^{i}$ shall be denoted, respectively, by 
$\partial_{\tau}$ and by $\partial_{i}$. }:
\begin{eqnarray}
(N^2 - N_{k} N^{k}) &=& (\alpha^2 - \beta_{k} \beta^{k}) (\partial_{\tau} T)^2 - 2 \beta_{i} \partial_{\tau} T \partial_{\tau}  Y^{i} - \overline{\gamma}_{ij} 
\partial_{\tau} Y^{i} \partial_{\tau} Y^{j},
\label{admin4a}\\
N_{i} &=& - (\alpha^2 - \beta_{k} \beta^{k}) \partial_{\tau} T \partial_{i} T + 2 \beta_{k}  \partial_{\tau} T \, \partial_{i} Y^{k}
+ \overline{\gamma}_{\ell k} \partial_{\tau} Y^{\ell} \partial_{i} Y^{k},
\label{admin4b}\\
\gamma_{ij} &=& - (\alpha^2 - \beta_{k} \beta^{k}) \partial_{j} T \partial_{i} T + 2 \beta_{k} \partial_{i} T \, \partial_{j}Y^{k} + \overline{\gamma}_{\ell k} \partial_{i} Y^{k} \partial_{j} Y^{\ell},
\label{admin4c} 
\end{eqnarray}
where the lapse function, the shift vectors and the 
spatial three metric in the transformed system have been denoted, respectively, by  $\alpha= \alpha(T,\vec{Y})$, $\beta_{i}= \beta_{i}(T,\vec{Y})$
and $\overline{\gamma}_{\ell k} = \overline{\gamma}_{\ell k}(T,\vec{Y})$; Eq. (\ref{adm1}) becomes, in the transformed frame, 
\begin{equation}
ds^2 = G_{\mu\nu}(T,\vec{Y})\,dY^{\mu} \,dY^{\nu} =\alpha^2 d T^2 - \overline{\gamma}_{ij} ( d Y^{i} + \beta^{i} d T)   ( d Y^{j} + \beta^{j} d T). 
\label{adm11}
\end{equation}
We shall often refer to the concept of non-linear gauge-invariant variables. This terminology refers to the possibility of finding 
specific quantities that do not depend on the choice of time hypersurfaces and that are exactly invariant for infinitesimal coordinate 
transformations in the perturbative regime. 

\renewcommand{\theequation}{3.\arabic{equation}}
\setcounter{equation}{0}
\section{Gravitating viscous fluids and gradient expansion}
\label{sec3}
Whenever dissipative effects are included both in the energy-momentum tensor and in the particle current the physical meaning of 
he four-velocity $u^{\mu}$ must be 
specified.  In the Eckart approach $u^{\mu}$ coincides with the velocity of particle transport. Conversely, in the Landau approach  the velocity $u^{\mu}$ coincides with the velocity of the energy transport defined by the $(0 i)$ component of the energy-momentum tensor giving the energy flux.  The Landau approach shall be privileged mainly for practical reasons.

\subsection{First and second viscosity in the Landau frame}
The total energy-momentum tensor of the problem is given as the perfect field contribution (characterized by 
a total pressure $p_{t}$ and a total energy density $\rho_{t}$) supplemented by the irreversible contribution: 
\begin{equation}
T_{\mu}^{\nu} = (p_{t} + \rho_{t}) u_{\mu} u^{\nu} - p_{t} \delta_{\mu}^{\nu} + {\mathcal T}_{\mu}^{\nu}(\eta,\xi),
\label{T1}
\end{equation}
where ${\mathcal T}_{\mu}^{\nu}(\eta,\xi)$ denotes the viscous energy momentum tensor depending on the 
first and second viscosities (i.e. $\eta$ and $\xi$):
\begin{equation}
{\mathcal T}_{\mu}^{\nu}(\eta,\xi) = 2 \eta \sigma_{\mu}^{\nu}   + \xi {\mathcal P}_{\mu}^{\nu} \nabla_{\alpha} u^{\alpha},\qquad \sigma_{\mu\nu} =
\frac{1}{2}  {\mathcal P}_{\mu}^{\gamma} \,{\mathcal P}_{\nu}^{\delta} \,{\mathcal W}_{\gamma\delta}.
\label{T2}
\end{equation}
As usual,  ${\mathcal P}_{\beta}^{\nu} = (\delta_{\beta}^{\nu} - u_{\beta} u^{\nu})$ and the tensor ${\mathcal W}_{\gamma\delta}$  appearing in Eq. (\ref{T2}) is defined as:
\begin{equation}
{\mathcal W}_{\gamma\delta} = \nabla_{\gamma} u_{\delta} + \nabla_{\delta} u_{\gamma} - \frac{2}{3} \, g_{\gamma\delta} \, \nabla_{\lambda} u^{\lambda}.
\label{T2a}
\end{equation}
The total particle current will be denoted by $j^{\mu} = n_{t} u^{\mu} + \nu^{\mu}$ (where $n_{t}$ is the total concentration of the fluid while  $\nu^{\mu}$ denotes the diffusion current). Note that $\nu^{\mu}$ will denote hereunder the relativistic thermal conduction four-vector. As we shall see explicitly 
$\nu^{\mu} u_{\mu} =0$ since $\nu_{\mu}$ can be written as 
$\nu_{\mu} = f(T, \rho_{t},p_{t} ) {\mathcal P}_{\mu}^{\alpha} \partial_{\alpha} \overline{\mu}$ where $\overline{\mu} = \mu/T$ and $\mu$ is the chemical potential.  

The viscous energy-momentum tensor at large-scales can be evaluated in the Landau-Lifshitz or  
in the Eckart frames. In the Eckart case the four-velocity $u_{\mu}$ appearing in Eq. (\ref{T1}) denotes 
the velocity of the particle transport. The total\footnote{In this paper we discuss the global evolution of large-scale curvature perturbations. We shall therefore deal with global quantities such as the total pressure, the total energy density of the system, the total viscosity of the fluid and so on. Of course the same analysis can be extended to the case where the fluid is composed by a number of fluids interacting among themselves as it happens, for instance, prior to photon decoupling.} particle current 
$j^{\mu}$ vanishes in the comoving frame.  Consequently the Eckart frame is fixed by requiring that $j^{\alpha} \, u_{\alpha} =0$ while ${\mathcal T}^{\mu\nu} u_{\nu} \neq 0$.  The Eckart approach \cite{visc3} (see also second paper of Ref. \cite{visc1}),  seems preferable when the concentration of radiation quanta exceeds the concentration of the other species.

Conversely, in the Landau-Lifshitz approach \cite{visc2} pure thermal conduction corresponds to an energy flux without particles: the four-velocity $u_{\mu}$ coincides with the velocity of the energy transport implying ${\mathcal T}^{\mu\nu} u_{\nu} =0$ . The two approaches are largely equivalent but the Landau-Lifshitz approach seems slightly more convenient, in the present context. 
Both the Eckart and the Landau approaches are suitable for the present class of problems where the typical scales are much larger than the mean free path. The second-order dissipative effects \cite{isr} become particularly relevant 
 in the collisions of heavy ions \cite{RHIC} where, however, not all the numerous second-order terms have been so far included in the available analytical and numerical discussions.
 
The covariant conservation of the total energy-momentum tensor (i.e. $\nabla_{\mu} T^{\mu\nu}=0$) can be 
projected along $u_{\nu}$ and along ${\mathcal P}_{\nu}^{\alpha}$; the two obtained equations together with the covariant conservation of the particle current are given hereunder:
\begin{eqnarray}
&& \nabla_{\mu}[ (p_{t} + \rho_{t}) u^{\mu}] - u_{\alpha} \partial^{\alpha} p_{t} + u_{\beta} \nabla_{\alpha} {\mathcal T}^{\alpha\beta} =0,
\label{c1}\\
&& (p_{t} + \rho_{t}) u^{\beta} \nabla_{\beta} u^{\alpha} - \partial^{\alpha}p_{t} + u^{\alpha} u_{\beta} \partial^{\beta} p_{t} + {\mathcal P}^{\alpha}_{\nu} \nabla_{\mu} {\mathcal T}^{\mu\nu} =0,
\label{c2}\\
&& \nabla_{\alpha}( n_{t} u^{\alpha} + \nu^{\alpha}) =0.
\label{c3}
\end{eqnarray}
Using then Eqs. (\ref{c1}) and (\ref{c3}) together  
with the first principle of thermodynamics, the evolution of the entropy\footnote{The explicit form of Eq. (\ref{entro}) has been obtained by trading the term $u_{\nu} \nabla_{\mu} {\mathcal T}^{\mu\nu}$ for 
$(\nabla_{\mu} u_{\nu}) {\mathcal T}^{\mu\nu}$ since, in the Landau frame, $\nabla_{\mu} ( u_{\nu} {\mathcal T}^{\mu\nu} ) = 0$.} can be easily derived:
\begin{equation}
\nabla_{\alpha} [ s u^{\alpha} - \overline{\mu} \nu^{\alpha} ] + \nu^{\alpha} \partial_{\alpha} \overline{\mu} = \nabla_{\alpha} u_{\beta} \, {\mathcal T}^{\alpha\beta}/T.
\label{entro}
\end{equation}
In Eq. (\ref{entro}), as already mentioned after Eq. (\ref{T2a}), $\overline{\mu} = \mu/T$ is the chemical potential rescaled through the temperature, $s$ is the entropy density and $\nu_{\alpha}$ is given by:
\begin{equation}
\nu_{\alpha} = \chi \biggl(\frac{n_{t} T}{\rho_{t} + p_{t}}\biggr)^2 \biggl[ \partial_{\alpha} \overline{\mu} - u_{\alpha} u^{\beta}\partial_{\beta} \overline{\mu} \biggr],
\label{nunu}
\end{equation}
where $\chi$ denotes the heat transfer coefficient. 
From the definition of the viscous energy-momentum tensor of Eqs. (\ref{T2}) and (\ref{T2a}) we can also explicitly write the term 
at the right hand side of Eq. (\ref{entro}) 
\begin{equation}
(\nabla_{\alpha} u_{\beta})\,{\mathcal T}^{\alpha\beta}/T = (\xi/T) (\nabla_{\alpha} u^{\alpha})^2 + 2 (\eta/T)\, \sigma_{\mu\nu} \,\sigma^{\mu\nu}.
\label{entro2}
\end{equation}
The adiabatic limit is recovered when the viscous contributions are neglected and the total entropy four-vector is conserved.
The right hand side of Eq. (\ref{entro2}) is positive semi-definite provided $\xi$ and $\eta$ are both positive semi-definite.

It is appropriate to remark, at this point, that the perfect fluid contribution is characterized by 
the barotropic index $w= p_{t}/\rho_{t}$ and by the related sound speed $c_{st}^2 = p_{t}^{\prime}/\rho^{\prime}$. 
In linear theory the fluctuations of the total pressure of the fluid are customarily decomposed into an adiabatic component supplemented by the entropic (or simply non-adiabatic) contributions $\delta p_{t} = c_{st}^2 \delta \rho_{t} +  \delta p_{\mathrm{nad}}$ (see, for instance, \cite{nonad} and the Eqs. (\ref{RPR})--(\ref{zetapr}) in section \ref{sec5}). This occurrence would correspond, at the level of the non-linear discussion, to the case $w \neq c_{st}^2$ where $w$ may be a space-time dependent function (see section \ref{sec4}).  Even if the conventional terminology might suggest otherwise, the non-adiabatic modes have nothing to do with the global viscosity of the system and may even arise in a globally inviscid fluid. This potential confusion of the standard terminology should be borne in mind to avoid unwanted misunderstandings. 

\subsection{Gradient expansion of the Einstein equations}
The Ricci tensor reported in Eqs. (\ref{ADM7}), (\ref{ADM8}) and (\ref{ADM9}) have been 
already written in a form where the terms containing spatial gradients are distinguished from all the other. 
The same criteria must be adopted when expressing the explicit components of the total energy-momentum tensor
so that, at the very end, we shall be able to write down the Einstein equations in their contracted form:
\begin{equation}
R_{\mu}^{\nu} = \ell_{P}^2 \biggl[ T_{\mu}^{\nu} - \frac{T}{2} \delta_{\mu}^{\nu} \biggr],\qquad \ell_{P} = \sqrt{8 \pi G},
\label{T6}
\end{equation}
where $T = T_{\alpha}^{\alpha}$ is the trace of the total energy-momentum tensor and must not be confused 
with the effective temperature of the fluid appearing in the previous subsection.  

The parameter counting the gradients 
can be formally indicated as the gradient itself in units of the trace of the extrinsic curvature, i.e.  as  $\lambda= \nabla/K(\tau,\vec{x})$. 
From Eq. (\ref{T1}) and bearing in mind the right hand side of Eq. (\ref{T6}) we have:
\begin{eqnarray}
T_{\mu\nu} - \frac{T}{2} g_{\mu\nu} &=& (\rho_{t} + P_{eff}) u_{\mu} u_{\nu} - \frac{\rho_{t} - P_{eff}}{2} g_{\mu\nu}  
\nonumber\\
&+& \eta\biggl\{ \nabla_{\mu} u_{\nu} + \nabla_{\nu} u_{\mu} - u^{\alpha} \biggl[ u_{\mu} \nabla_{\alpha} u_{\nu} + u_{\nu} \nabla_{\alpha} u_{\mu} \biggr]
- \frac{2}{3} {\mathcal P}_{\mu\nu} \nabla_{\lambda} u^{\lambda} \biggr\}.
\label{EM1}
\end{eqnarray}
The trace $T$ has been expressed in terms of the effective pressure $P_{eff}$ defined, in our case, as:
\begin{equation}
T = T_{\alpha}^{\alpha} = \rho_{t} - 3 P_{eff}, \qquad P_{eff} = p_{t} - \xi \nabla_{\alpha} u^{\alpha}.
\end{equation}
The term $\nabla_{\alpha} u^{\alpha}$ can be easily expanded in spatial gradients and the result is:
\begin{equation}
\nabla_{\alpha} u^{\alpha} = - K   -  \frac{1}{N \sqrt{\gamma}} \partial_{k} [ N 
\sqrt{\gamma} u^{k}] + \frac{\partial_{\tau} u^2}{2 N} + {\mathcal O}(\lambda^3).
\end{equation}
From Eqs. (\ref{ADM7}), (\ref{ADM8}) and (\ref{ADM9}), the terms containing the shift 
vectors turn out to be ${\mathcal O}(\lambda)$ (see also \cite{salope,shell}). It seems therefore appropriate to select the gauge $N_{i} =0$ where the coordinate observers coincide with normal observers. Equivalent choices, for the present purposes, include the coordinate system where the expansion is uniform (i.e. $K = K(\tau)$, in our notations) or the gauge where the energy density is uniform.

The full expression of the various components of Eq. (\ref{EM1}) are necessarily lengthy so we shall just exemplify 
the $(00)$ component and then mention the leading order results for the other components. The $(00)$ component of Eq. (\ref{EM1}) 
is given by:
\begin{equation}
T_{00} - \frac{T}{2} g_{00} = \frac{N^2}{2} ( \rho_{t} + 3 P_{eff}) + N^2 u^2 ( \rho_{t} + P_{eff}) - 2 \eta N u^2 {\mathcal F}(N,\gamma_{ij}, u_{k}),
\label{inter1}
\end{equation}
where  ${\mathcal F}(N,\gamma_{ij}, u_{k})$ is defined as:
\begin{eqnarray}
 {\mathcal F}(N,\gamma_{ij}, u_{k}) &=& \biggl\{ \partial_{\tau}[\sqrt{1 + u^2}] - u_{i} \partial^{i} N - 
N \sqrt{ 1 + u^2} \biggl[ u^{k} \partial_{k} \sqrt{1 + u^2} + u^{k} u^{j} K_{k j} \biggr] 
\nonumber\\
&-& \frac{1}{3 \sqrt{\gamma}} [ \partial_{\tau} ( \sqrt{\gamma} \sqrt{1 + u^2}) - \partial_{k}(N \sqrt{\gamma} u^{k})]\biggr\}.
\label{inter1a}
\end{eqnarray}
Except for the first term at the right hand side, all the remaining contributions are ${\mathcal O}(\lambda^2)$ since they contain, at 
least, two spatial gradients. From the lowest-order form of the momentum constraint (see below Eqs. (\ref{inter2}) and (\ref{T8})) the leading contribution 
of the spatial part of the velocity is clearly ${\mathcal O}(\lambda)$ so that $u^2 = \gamma_{ij} u^{i} u^{j} = {\mathcal O}(\lambda^2)$. 

The same procedure outlined in the case of the $(00)$ component of the energy-momentum tensor can be discussed for the remaining components 
and the leading order results are:
\begin{eqnarray}
T_{i}^{0} &=& (\rho_{t} + P_{eff}) u_{i} \, u^{0} + 2 \eta u_{k} u^{0} \overline{K}_{i}^{k}  + {\mathcal O}(\lambda^3),
\label{inter2}\\
T_{i}^{j} - \frac{T}{2} \delta_{i}^{j} &=& -\frac{1}{2} ( \rho_{t} - P_{eff}) \delta_{i}^{j} - 2 \eta \overline{K}_{i}^{j}  + {\mathcal O}(\lambda^2),
\label{inter3}
\end{eqnarray} 
where the traceless part of the extrinsic curvature $\overline{K}_{i}^{j} = K_{i}^{j} - \delta_{i}^{j} K/3$ has been explicitly introduced.

Using the results of Eqs. (\ref{inter1}), (\ref{inter2}) and (\ref{inter3}) together with  Eqs. (\ref{ADM7}), (\ref{ADM8}) and (\ref{ADM9}),
Eq. (\ref{T6}) will become:
\begin{eqnarray}
&& \frac{1}{N}\partial_{\tau} K - \mathrm{Tr} K^2 = \frac{\ell_{P}^2}{2}  ( \rho_{t} + 3 P_{eff}),
\label{T7}\\
&& \nabla_{i} K - \nabla_{k} K^{k}_{i} = N \ell_{P}^2 \biggl[ ( \rho_{t} + P_{eff}) u_{i} u^{0} + 2\eta \overline{K}_{i}^{j} u_{j} u^{0} \biggr],
\label{T8}\\
&& \frac{1}{N}\partial_{\tau} K_{i}^{j} -  K K_{i}^{j} -  r_{i}^{j} = \ell_{P}^2 \biggl[ \frac{(P_{eff} - \rho_{t})}{2} \delta_{i}^{j} - 2 \eta \overline{K}_{i}^{j} + \Pi_{i}^{j} \biggr],
\label{T9}
\end{eqnarray}
where $\Pi_{i}^{j}$ denotes the  anisotropic stress which is by definition a traceless rank-two tensor in three dimensions. We have chosen 
to keep generic the form of $\Pi_{i}^{j}$ since it may contain all the potential sources of anisotropic stress 
not necessarily connected to the the fluid sector such as scalar fields or even gauge fields. In all cases $\Pi_{i}^{j}$ contains at least two 
spatial gradients. The anisotropic stress and the curvature $r_{i}^{j}$ are of higher order in the gradients but have been kept for the benefit of the forthcoming discussion aimed at showing the the traceless part of the extrinsic curvature is of higher order in the gradients and it the only 
component affected by the presence of shear viscosity.

Finally, the evolution equations stemming from the covariant conservation 
of the energy-momentum tensor and of the particle current  of Eqs. (\ref{c1}), (\ref{c2}) and (\ref{c3}) are given by 
\begin{eqnarray}
&& \frac{1}{N} \partial_{\tau} \rho_{t} -  K ( \rho_{t} + P_{eff}) =0
\label{EN}\\
&& \frac{1}{N}\partial_{\tau} u^{i} + u^{k} \biggl[ \frac{\partial_{\tau} P_{eff}}{N(\rho_{t} + P_{eff})} \delta_{k}^{i} - 2  K_{k}^{i} \biggr] = \frac{\partial^{i} N}{N} - \frac{\partial^{i} P_{eff}}{\rho + P_{eff}},
\label{VEL}\\
&& \frac{1}{N} \partial_{\tau} n_{t} - K n_{t} =0.
\label{dens}
\end{eqnarray}
The absence of the dissipative coefficients arising in the diffusion current is justified since these terms are of higher 
order in the spatial gradients, as it can be easily appreciated from Eq. (\ref{nunu}).

\subsection{Decoupling of the shear contribution}
From Eq. (\ref{T9}) the shear contribution only affects the evolution of the traceless part of the extrinsic curvature 
$\overline{K}_{i}^{j}$ and does not enter the deceleration parameter whose sign is solely determined by the bulk viscosity 
coefficient. 

Indeed, after taking the  the traceless part of Eq. (\ref{T9}) the following equation is obtained:
\begin{equation} 
\partial_{\tau} \overline{K}_{i}^{j} - N K \overline{K}_{i}^{j}  = - 2 \eta N \ell_{P}^2  \overline{K}_{i}^{j} + N \ell_{P}^2 \Pi_{i}^{j} + N \overline{r}_{i}^{j}
\label{S1}
\end{equation}
where $\overline{r}_{i}^{j} = r_{i}^{j} - \delta_{i}^{j} \, r/3$ is the traceless part of the spatial curvature. In the general situation where $\eta(\tau, \vec{x})$ Eq. (\ref{S1}) implies
\begin{eqnarray}
\overline{K}_{i}^{j}(\tau,\vec{x}) &=& \frac{\sqrt{\gamma(\tau_{*}, \vec{x})}}{\sqrt{\gamma(\tau,\vec{x})}} \, \overline{K}_{i}^{j}(\tau_{*},\vec{x}) e^{- 2 {\mathcal A}(\tau_{*}, \tau, \vec{x})}
\nonumber\\
&+& \frac{\ell_{P}^2}{\sqrt{\gamma(\tau,\vec{x})}} \int_{\tau_{*}}^{\tau} d\tau^{\prime\prime} \sqrt{\gamma(\tau^{\prime\prime},\vec{x})}
N(\tau^{\prime\prime},\vec{x})\, e^{ - 2 {\mathcal A}(\tau^{\prime\prime},\tau,\vec{x})}\, \Pi_{i}^{j}(\tau^{\prime\prime}, \vec{x}) d\tau^{\prime\prime}
\nonumber\\
&+& \frac{1}{\sqrt{\gamma(\tau,\vec{x})}} \int_{\tau_{*}}^{\tau} d\tau^{\prime\prime} \sqrt{\gamma(\tau^{\prime\prime},\vec{x})}
N(\tau^{\prime\prime},\vec{x})\, e^{ - 2 {\mathcal A}(\tau^{\prime\prime},\tau,\vec{x})}\, \overline{r}_{i}^{j}(\tau^{\prime\prime}, \vec{x}) d\tau^{\prime\prime}
\label{S2}
\end{eqnarray}
where $\gamma = \mathrm{det}(\gamma_{ij})$. In Eq. (\ref{S2}) $\tau_{*}= \tau_{*}(\vec{x})$ denotes some arbitrary integration time while for two generic times $\tau_{1}$ and $\tau_{2}$, 
${\mathcal A}(\tau_{1},\tau_{2},\vec{x})$ is defined as: 
\begin{equation}
{\mathcal A}(\tau_{1},\tau_{2},\vec{x}) = \ell_{P}^2 \int_{\tau_1}^{\tau_{2}} \eta(\tau^{\prime},\vec{x}) \, N(\tau^{\prime},\vec{x}) \, d\tau^{\prime}.
\label{S3}
\end{equation}
Equations (\ref{S2}) and (\ref{S3}) show that the traceless part of the extrinsic curvature 
is determined by the anisotropic stress and by the traceless part of the intrinsic curvature. Both 
quantities are of higher order in the gradient expansion\footnote{The higher order terms in the gradient expansion 
can be computed by following iterative methods where the spatial geometry is reconstructed, order by order, starting from a 
seed metric that do not contain any spatial gradient \cite{salope,tomita} but this is not our primary goal in this investigation.}. 
Equation (\ref{S3}) shows that the shear viscosity suppresses the traceless part of the extrinsic curvature even further in comparison with the case $\eta \to 0$. The features of the damping  are determined by the explicit expression of $\eta$. For a system dominated by radiation $\eta \sim \ell_{mfp} \, \rho_{t}$ 
where $\ell_{mfp}$ denotes the mean free path (for instance the Thomson mean free path prior to photon decoupling). In this case 
${\mathcal A} \simeq K \ell_{mfp} \ll 1$. In more general terms, however, $\eta$ can depend on $\rho_{t}$ on the trace of the extrinsic curvature, on the total particle concentration and in all these cases $\overline{K}_{ij}$ may even be much smaller than 
the anisotropic stress. 

Since $\eta$ decouples from the trace of the extrinsic curvature, it does not 
contribute to the inhomogeneous generalization of the deceleration parameter.  For the sake of comparison with the fully homogenous case we choose Gaussian normal coordinates and set $N=1$; in this situation Eq. (\ref{T7}) can be written as:
\begin{equation}
q(t,\vec{x}) \mathrm{Tr} K^2 = \ell_{P}^2 \biggl[ (\rho_{t} + P_{eff}) u_{0} u^{0} + \frac{P_{eff} - \rho_{t}}{2}\biggr] - 2 \eta u^2 {\mathcal F}(1,\gamma_{ij},u_{k}),
\label{00cont}
\end{equation}
where $q(\vec{x},t) = -1 + \dot{K}/{\rm Tr} K^2$ is the inhomogeneous generalisation of the deceleration parameter\footnote{In the homogeneous and isotropic limit, $ \gamma_{ij} = a^2(t) 
\delta_{ij}$, $K_{i}^{j} = - H \delta_{i}^{j}$ and, as expected, $q(t) \to - \ddot{a} a/ \dot{a}^2$.}  and the overdot denotes the derivative with 
respect to the cosmic time coordinate $t$ which coincides with $\tau$ in the case $N=1$. The function ${\mathcal F}(1,\gamma_{ij},u_{k})$ (defined in Eq. (\ref{inter1a})) accounts for the higher-order corrections.
In general terms ${\rm Tr} K^2 \geq K^2/3 \geq 0$,
where the sign of equality (in the first relation) is reached in 
the isotropic limit. Since $\gamma^{ij}$ is always positive semi-definite, it is
also clear that $u_{0}\,u^{0} = 1 + \gamma^{ij} u_{i} u_{j} \geq 1$.
From Eq. (\ref{00cont}) it also follows
that $q(t,\vec{x})$ is always positive semi-definite as long as $(\rho + 3 P_{eff}) \geq 0$. This means that the sign of the generalized 
deceleration parameter only depends on $P_{eff}$ (and hence on the bulk viscosity) while the shear viscosity 
does not play any role. According to Eq. (\ref{00cont}) the correction of the bulk viscosity only arises to second order in the gradient expansion where, however, the bulk viscosity also contributes through the term $(\rho + P_{eff}) u^2$ implicitly contained in $ (\rho_{t} + P_{eff}) u_{0} u^{0}$.

\renewcommand{\theequation}{4.\arabic{equation}}
\setcounter{equation}{0}
\section{Gauge invariant variables and their evolution}
\label{sec4}
\subsection{Generalities}
Under Eq. (\ref{adm2}) the energy-momentum tensor  transforms as
\begin{equation}
T_{\mu\nu}(\tau,\vec{x}) \to \overline{T}_{\mu\nu}(T, \vec{Y}) = \biggl(\frac{\partial x^{\alpha}}{\partial Y^{\mu}} \biggr) \biggl(\frac{\partial x^{\beta}}{\partial Y^{\nu}} \biggr) T_{\alpha\beta}(\tau,\vec{x}),
\label{TTtr}
\end{equation}
where $T$ is the transformed time coordinate and will not be confused with the trace of the energy-momentum tensor.
Equations (\ref{admin4a})--(\ref{admin4b}) and (\ref{admin4c}) hold for the metric; similar expressions hold for the transformed components of the energy-momentum tensor. The analog of Eq. (\ref{admin4a}) will be reported to fix the notations
\begin{equation}
T_{00}(\tau,\vec{x}) = (\partial_{\tau} T)^2 \overline{T}_{00}(T,\vec{Y}) + 2 (\partial_{\tau} T) (\partial_{\tau} Y^{i}) \overline{T}_{0i}(T,\vec{Y})
+(\partial_{\tau} Y^{i}) (\partial_{\tau} Y^{j}) \overline{T}_{ij}(T,\vec{Y}).
\label{TTtr2}
\end{equation}
The explicit expressions of the $(0i)$ and $(ij)$ components can be easily written in terms of the notations of Eq. (\ref{TTtr2}) and will be 
employed below.

The coordinate transformation must preserve the order of the gradient expansion. 
This implies, from Eq. (\ref{admin4b}), that $\beta_{i} = N_{i} =0$ and the coordinate transformation demands:
\begin{equation}
 \alpha^2 \partial_{\tau} T \partial_{i} T = \overline{\gamma}_{\ell k} \partial_{\tau} Y^{\ell} \partial_{i} Y^{k}.
 \label{first}
\end{equation}
The transformations preserving the order of the gradient expansion \cite{salope} can be written as follows:
\begin{equation}
\tau \to T= T(\tau, \vec{x}), \qquad x^{k} \to Y^{k}(\tau,\vec{x}) = f^{k}(\tau,\vec{x}) + F^{k}(\tau,\vec{x}).
\label{trans}
\end{equation}
By construction the function $f^{i}(\vec{x},\tau)$ does not contain any gradient while $F^{i}(\tau,\vec{x})$ contains at least one spatial gradient;
$f^{k}(\tau, \vec{x})$ can then be parametrized as $f^{k}(\tau,\vec{x}) = x^{k} g(\tau,r)$ where $r = \sqrt{x_{i} x^{i}}$. Since in the transformation 
all the spatial gradients of $g(\tau,r)$ will automatically contribute to $F^{k}(\tau, \vec{x})$, the  effect of $g(\tau,r)$ boils down to a redefinition of $\alpha$ in the transformed frame. For this reason and for the sake of simplicity we shall set $g(r,\tau)=1$. 

Equations (\ref{admin4a}) and (\ref{admin4c}), thanks to Eq. (\ref{first}), will then lead, respectively, to the following pair of conditions:
\begin{eqnarray}
&& N^2 = \alpha^2 (\partial_{\tau} T)^2 - \overline{\gamma}_{ij} \partial_{\tau} Y^{i} \partial_{\tau} Y^{j}, 
\label{second}\\
&& \gamma_{ij} = - \alpha^2 \partial_{i}T \partial_{j} T + \overline{\gamma}_{k\ell} \partial_{i} Y^{k} \, \partial_{j} Y^{\ell}.
\label{third}
\end{eqnarray}
Recalling the explicit form of Eq. (\ref{trans}), to lowest order in the spatial gradients, Eq. (\ref{third}) implies 
that $\gamma_{ij}(\tau, \vec{x})= \gamma_{ij}(T, \vec{Y})$ while Eqs. (\ref{first}) and (\ref{second}) determine 
the explicit form of $F^{k}(T,\vec{Y})$; the explicit results are:
\begin{equation}
F^{k}(T,\vec{Y}) = \int d T \, N^2(\tau) \frac{\partial^{k} T}{(\partial_{\tau} T)^2}, \qquad \gamma_{ij}(\tau, \vec{x})= \gamma_{ij}(T, \vec{Y}).
\label{fourth}
\end{equation}
Equations (\ref{trans}) and  (\ref{fourth}) can be inserted into the various components of Eq. (\ref{TTtr}) to obtain the transformation 
properties of the pressure, of the energy density and of the velocity:
\begin{eqnarray}
&&  \rho(\tau,\vec{x})=\overline{\rho}(T,\vec{Y}) , \qquad P_{eff}(\tau,\vec{x})= \overline{P}_{eff}(T,\vec{Y}),
\nonumber\\
&& \qquad u_{i} = \overline{u}_{i} + \alpha \partial_{i}T, \qquad 
N = \alpha \,\partial_{\tau} T.
\label{fifth}
\end{eqnarray}

\subsection{Non-linear curvature inhomogeneities}
In the viscous case the non-linear generalization of the curvature perturbations on comoving orthogonal hypersurfaces 
is:
\begin{equation}
{\mathcal R}_{i}(\tau,\vec{x}) = \frac{1}{3} \nabla_{i} [\ln{(\sqrt{\gamma})}] - \frac{1}{3N} \partial_{\tau}[ \ln{(\sqrt{\gamma})}] \,\, u_{i}.
\label{defR}
\end{equation}
Using the transformation properties defined by Eqs. (\ref{first}), (\ref{second}), (\ref{third}) and (\ref{fifth}) we have that ${\mathcal R}_{i}$ transforms as:
\begin{eqnarray}
{\mathcal R}_{i}(\tau, \vec{x}) &\to& \overline{{\mathcal R}}_{i}(T, \vec{Y}) = \frac{1}{3} 
\frac{\partial [\ln{(\sqrt{\overline{\gamma}})}]}{\partial Y^{j}}
 \frac{\partial Y^{j}}{\partial x^{i}} + \frac{1}{3} \frac{\partial[ \ln{(\sqrt{\overline{\gamma}})}]}{\partial T} \frac{\partial T}{\partial x^{i}} 
\nonumber\\
&-& \frac{1}{3 \alpha} \biggl(\overline{u}_{i} + \alpha \frac{\partial T}{\partial x^{i}} \biggr)   \frac{\partial [\ln{(\sqrt{\overline{\gamma}})}]}{\partial T};
\label{transR1}
\end{eqnarray}
since the two intermediate terms simplify in Eq. (\ref{transR1}), we have that the curvature inhomogeneities are invariant i.e. ${\mathcal R}_{i}(\tau,\vec{x}) = \overline{{\mathcal R}}_{i}(T,\vec{Y})$.  

Using the same strategy applied in the case of Eq. (\ref{defR}), the non-linear generalization  of the density contrast on uniform curvature hypersurfaces becomes\footnote{In linear theory the density contrast on uniform curvature hypersurfaces is invariant under infinitesimal coordinate transformations. Since, by definition, it has the same value in different gauges it can be also interpreted as the curvature perturbation on the hypersurfaces where the energy density is unperturbed. These two physical interpretations are relevant when discussing the so-called 
$\delta {\mathcal N}$ formalism (see section \ref{sec5}). }:
\begin{equation}
\zeta_{i}(\tau,\vec{x}) =  \frac{1}{3} \nabla_{i}[ \ln{(\sqrt{\gamma})} ]+ \frac{\nabla_{i} \rho}{ 3(\rho + P_{eff})}.
\label{defZ}
\end{equation}
The same analysis leading to Eq. (\ref{transR1}) can be performed in the case of the $\zeta_{i}(\tau,\vec{x})$:
\begin{eqnarray}
\zeta_{i}(\tau, \vec{x}) &\to& \overline{\zeta}_{i}(T, \vec{Y}) =\frac{1}{3} 
\frac{\partial [\ln{(\sqrt{\overline{\gamma}})}]}{\partial Y^{j}}
 \frac{\partial Y^{j}}{\partial x^{i}} + \frac{1}{3} \frac{\partial[ \ln{(\sqrt{\overline{\gamma}})}]}{\partial T} \frac{\partial T}{\partial x^{i}} 
\nonumber\\
&+& \frac{1}{3 (\overline{\rho}_{t} + \overline{P}_{eff})} \,\frac{\partial \overline{\rho}}{\partial Y^{j}} \frac{\partial Y^{j}}{\partial x^{i}} + \frac{1}{3 (\overline{\rho}_{t}+ \overline{P}_{eff})} \frac{\partial \overline{\rho}}{\partial T} \frac{\partial T}{\partial x^{i}}.
\label{transzeta1}
\end{eqnarray}
 Equation (\ref{EN}) 
(stemming from the covariant conservation of the total energy-momentum tensor in the transformed frame) implies that the derivative
of $\overline{\rho}_{t}$ with respect to $T$ equals $\alpha \overline{K} ( \overline{\rho}_{t} + 
\overline{P}_{eff})$. Thus the second and fourth terms at the right hand side of Eq. (\ref{transzeta1}) cancel since $\alpha \overline{K}= - 
\partial \ln{\sqrt{\overline{\gamma}}}/\partial T$. As in the case of Eq. (\ref{transR1}) the invariance of $\zeta_{i}$ is manifest
since $\zeta_{i}(\tau,\vec{x}) = \overline{\zeta}_{i}(T,\vec{Y})$.

Since, in the general situation, the bulk viscosity coefficient is a space-time scalar function its derivative 
transforms non-trivially under Eq. (\ref{trans}):
\begin{equation}
\frac{\partial \xi }{\partial x^{i}}= \frac{\partial \overline{\xi}}{\partial Y^{i}} + \partial_{T} \overline{\xi} \partial_{i} T, \qquad \partial_{\tau} \xi = \frac{\partial \overline{\xi}}{\partial T} \partial_{\tau} T.
\label{Z2}
\end{equation}
Thanks to Eq. (\ref{Z2}) we can obtain a further non-linear variable invariant under Eqs. (\ref{first})--(\ref{second}) and (\ref{third})--(\ref{fifth}) that has no analogue in the inviscid case:
\begin{equation}
{\mathcal Z}_{i}(\tau, \vec{x}) = \frac{1}{3} \nabla_{i} [\ln{(\sqrt{\gamma})}]  + \frac{K N}{3} \frac{\partial_{i} \xi}{\partial_{\tau}\xi}.
\label{Z1}
\end{equation}
From the gauge-transformed expression of Eq. (\ref{Z1}), using  Eqs. (\ref{second}) and (\ref{third}) the gauge-invariance 
of ${\mathcal Z}_{i}(\tau, \vec{x})$ is easily demonstrated. 

Let us finally mention, for the sake of comparison, that the general form of Eq. (\ref{transR1}) can be used to recover the well known results obtainable in the case of the single scalar field \cite{salope}. 
The $(0i)$ component of the energy-momentum tensor 
of a minimally coupled scalar field $T_{i}^{0}(\varphi)=\partial_{i} \varphi \partial_{\tau}\varphi/N^2$ implies that $u_{i} = N \nabla_{i}\varphi /(\partial_{\tau}\varphi)$. 
Therefore Eq. (\ref{defR}) implies 
\begin{equation}
{\mathcal R}_{i}(\tau,\vec{x}) = \frac{1}{3} \nabla_{i}[\ln {(\sqrt{\gamma})}] + \frac{K N}{3} \frac{\nabla_{i} \varphi}{\partial_{\tau}\varphi},
\label{conv1}
\end{equation}
which is also invariant since, from Eq. (\ref{fifth}), the transformation of
$\varphi$ will be given by:
\begin{equation}
\partial_{\tau} \varphi= \partial_{T} \overline{\varphi} (\partial_{\tau} T),\qquad \frac{\partial \varphi}{\partial x^{i}} = \frac{\partial \overline{\varphi}}{\partial Y^{i}} + \partial_{T} \overline{\varphi} \partial_{i} T.
\label{conv2}
\end{equation}
Equations (\ref{conv1}) and (\ref{conv2}) reproduce the standard results of Refs. \cite{salope,shell}.

\subsection{Evolution of the gauge-invariant variables}
According to the momentum constraint of Eq. (\ref{T8}) the combination $K u_{i}$ appearing in Eq. (\ref{defR}) is expressible in terms of the gradients of the extrinsic curvature as:
\begin{equation}
K u_{i} = \frac{1}{3 \ell_{P}^2 ( \rho_{t}+ P_{eff})} \biggl[ \partial_{i} K^2 - 3K \nabla_{k} \overline{K}_{i}^{k}\biggr].
\label{A1}
\end{equation}
In Eq. (\ref{A1}) and in the forthcoming discussion we shall keep the dependence on the 
traceless part of the extrinsic curvature just to keep track of the difference between ${\mathcal R}_{i}$ and $\zeta_{i}$. 
From Eq. (\ref{T7}) and from the trace of Eq. (\ref{T9}) we obtain the following pair of equations:
\begin{equation}
2 \ell_{P}^2 \rho_{t} = K^2 - \mathrm{Tr}K^2,\qquad 3 N \ell_{P}^2 (\rho_{t} + P_{eff}) = 2 \partial_{\tau}K - 3 N \mathrm{Tr}K^2 + N K^2, 
\label{A2}
\end{equation}
implying that Eq. (\ref{A1}) can be finally expressed as:
\begin{equation}
K u_{i} = \frac{\partial_{i} \rho_{t}}{ \rho_{t} + P_{eff}} + {\mathcal G}_{i},\qquad  {\mathcal G}_{i}= \frac{1}{6 \ell_{P}^2( \rho_{t} + P_{eff})} \biggl\{ 3\partial_{i}[ \mathrm{Tr} \overline{K}^2] - 6 K\, \nabla_{k} \overline{K}^{k}_{i} \biggr\}.
\label{A3}
\end{equation}
After inserting Eq. (\ref{A3})  into Eq. (\ref{defR}), the partial time derivative of ${\mathcal R}_{i}$ becomes:
\begin{equation}
\partial_{\tau} {\mathcal R}_{i} = - \frac{1}{3} \partial_{i}( N K) + \frac{1}{3} \partial_{\tau} \biggl(\frac{\partial_{i} \rho_{t} }{\rho_{t} + P_{eff}}\biggr) + \frac{\partial_{\tau} {\mathcal G}_{i}}{3}.
\label{A3a}
\end{equation}
To leading order in the spatial gradients, the same kind of evolution equation reported in Eq. (\ref{A3a}) is derivable for 
$\zeta_{i}$ starting directly from the definition Eq. (\ref{defZ}) and using Eq. (\ref{A2}). The leading terms of the evolution 
equation will be the same and the rationale for this occurrence is that 
$\zeta_{i}$ and ${\mathcal R}_{i}$ differ by terms that are of higher order in the gradient expansion, i.e.
$6 \ell_{P}^2 \,(\zeta_{i} - {\mathcal R}_{i}) = -  [ 3 \partial_{i}( \mathrm{Tr}\overline{K}^2) - 6 K \nabla_{k} \overline{K}^{k}_{i}]$.

Inserting now Eq. (\ref{EN}) into the first term at the right hand side of Eq. (\ref{A3a}) we arrive at the following result:
\begin{equation}
\partial_{\tau} {\mathcal R}_{i} = \frac{1}{3} \partial_{\tau} \biggl(\frac{\partial_{i}\rho_{t}}{\rho_{t} + P_{eff}}\biggr) - \frac{1}{3} \partial_{i} \biggl(\frac{\partial_{\tau}\rho_{t}}{\rho_{t} + P_{eff}}\biggr)
+ \frac{\partial_{\tau} {\mathcal G}_{i}}{3}.
\label{A4}
\end{equation}
The third term at the right hand side of Eq. (\ref{A4}) will now be dropped since it is of higher order in the gradients. Equation (\ref{A4})
 can be expressed in a physically more significant form by separating the viscous contributions from the 
conventional non-adiabatic terms that normally appear even in the absence of irreversible contributions:
\begin{equation}
  \partial_{\tau} {\mathcal R}_{i} = {\mathcal S}_{nad}(\tau,\vec{x}) + {\mathcal S}_{viscous}(\tau,\vec{x}),
\label{A4a}
\end{equation}
where the two source terms are given, respectively, by
\begin{eqnarray}
{\mathcal S}_{nad}(\tau, \vec{x}) &=& \frac{ K N}{ 3 (\rho_{t} + P_{eff})} \bigl( \partial_{i} p_{t} - c_{st}^2 \partial_{i} \rho_{t}\bigr),
\label{A4b}\\
{\mathcal S}_{viscous}(\tau, \vec{x}) &=& \frac{K}{3 ( \rho_{t} + P_{eff})^2} \bigl[ (\partial_{\tau} \rho_{t}) \partial_{i} \xi - \partial_{i} \rho_{t} (\partial_{\tau} \xi)\bigr] 
\nonumber\\
&+& \frac{\xi}{3 (\rho_{t} + P_{eff})^2} \bigl[(\partial_{\tau} \rho_{t})\partial_{i} K - 
(\partial_{i} \rho_{t})\partial_{\tau} K \bigr].
\label{A4c}
\end{eqnarray}
As already mentioned, the total sound speed is $c_{st}^2 = \partial_{\tau} p_{t}/\partial_{\tau}\rho_{t}$. Let us consider first ${\mathcal S}_{nad}(\tau,\vec{x})$ and show 
that it is nothing but the standard adiabatic contribution. Broadly speaking the barotropic index is a space-time 
function, i.e. $w(\tau,\vec{x}) = p_{t}/\rho_{t}$ and Eq. (\ref{A4b}) implies: 
\begin{equation}
{\mathcal S}_{nad}(\tau, \vec{x}) = \frac{K N \rho_{t}}{3 ( \rho_{t} + P_{eff})} \partial_{i} w - \frac{\rho_{t} \partial_{i} \rho_{t}}{3 ( \rho_{t} + P_{eff})^2} \partial_{\tau} w.
\label{A4d}
\end{equation}
In linear theory when $w$ is a space-time constant the sound speed equals $\sqrt{w}$ and the non-adiabatic 
contribution is absent. In this situation ${\mathcal S}_{nad}(\tau, \vec{x}) \to 0$, as expected.

According to Eqs. (\ref{A4a}), (\ref{A4b}) and (\ref{A4c}) the relativistic viscous fluids lead to a source term implying that ${\mathcal R}_{i}$ is not constant in
general terms. This may happen in various situations where, for instance, the viscosity coefficients depend in time and in space.
For instance, across the matter-radiation transition the shear viscosity coefficient $\eta$  determines the optical depth, the Silk damping scale and, ultimately, the shape of the visibility function \cite{peebles,visibility}. The scaling properties of $\eta$ and $\xi$ can be expressed
$\xi/\eta \simeq \biggl( \frac{1}{3}- c_{\mathrm{st}}^2 \biggr)^{2 q}$ for $q\geq 1$. Across the matter-radiation transition the sound speed of the plasma interpolates between $1/\sqrt{3}$ and $0$. Adopting the viewpoint of linear theory (see section \ref{sec5}) and separating the background from the fluctuations the sound speed can be computed as $c_{st}^2 = 4/[3 ( 4 + 3 \alpha)]$ where $\alpha= a/a_{eq}$ denotes the scale factor normalized at equality. This dependence in $\xi$ implies the generation of non-adiabatic modes (see, in particular, the last paper of \cite{nonad}).

\subsection{Conservation of curvature perturbations}

If $c_{st}^2 \neq w$  (or if $w$ is a space-time function) the curvature 
perturbations are non conserved  even in 
the limit $\xi\to 0$, and this is nothing but the standard situation of the conventional non-adiabatic modes. 
To exclude all the potential sources that could make 
${\mathcal R}_{i}$  time dependent besides the ones we ought to investigate specifically, namely the relativistic viscous contributions we shall posit that $w$ is constant and that $ c_{st} = \sqrt{w}$. This choice implies, according to Eq. (\ref{A4b}), that ${\mathcal S}_{nad}=0$.

The only contribution remaining at the right hand side of Eq. (\ref{A4a}) is the one coming from ${\mathcal S}_{viscous}$. Furthermore, by focussing on Eq. (\ref{A4c}) we see that the second term (proportional to $\xi$) can be 
rewritten as 
\begin{equation}
\frac{\xi}{3 (\rho_{t} + P_{eff})^2} \bigl[(\partial_{\tau} \rho_{t})\partial_{i} K - 
(\partial_{i} \rho_{t})\partial_{\tau} K \bigr] = \frac{\xi N}{6 (\rho_{t} + P_{eff})} \partial_{i}\bigl( K^2 - 3 \ell_{P}^2 \rho_{t}),
\label{van1}
\end{equation}
but the term at the right hand side vanishes because of the first of Eq. (\ref{A2}); in fact, 
$\mathrm{Tr}K^2 = K^2/9 + {\mathrm Tr} \overline{K}^2$ and ${\mathrm Tr}\overline{K}^2$ is of higher order being proportional 
to the square of the total anisotropic stress. We stress that the result of Eq. (\ref{van1}) holds non-perturbatively; it does not assume a
separation between the background space-time and its perturbative fluctuations.
Equation (\ref{A4a}) becomes then:
\begin{equation}
 \partial_{\tau} {\mathcal R}_{i} = \frac{K}{3 ( \rho_{t} + P_{eff})^2} 
 \bigl[ (\partial_{\tau} \rho_{t}) \partial_{i} \xi - \partial_{i} \rho_{t} (\partial_{\tau} \xi)\bigr]. 
\label{A5}
\end{equation} 
If the source term in Eq. (\ref{A5}) vanishes the curvature inhomogeneities will be conserved and the equations of motion will enjoy a further 
symmetry since ${\mathcal R}_{i}(\tau,\vec{x})$ can be shifted by a a term constant in time (but not in space). 
The bulk viscosity $\xi$ can depend, in principle, on five  quantities, namely  $K$,  $\mathrm{Tr}K^2$, 
  $\rho_{t}$, $p_{t}$ and $n_{t}$. The dependence on $p_{t}$ can be traded for $\rho_{t}$ since $p_{t} = w \rho_{t}$ with constant $w$. 
 Since the traceless part of the extrinsic curvature 
is of higher order we can also drop the dependence on $\mathrm{Tr}K^2$ that coincides, to leading order, with  $K^2/9$.
Equation (\ref{A2}) can be finally used to relate $K^2$ and $\rho_{t}$. 
To lowest order in the gradient expansion we have therefore 
only two qualitatively different cases: $\xi = \xi(\rho_{t})$ and $\xi=\xi(\rho_{t},n_{t})$; 
the case $\xi=\xi(n_{t})$ is indeed the same as  the one where $\xi(\rho_{t},n_{t})$. 

When $\xi = \xi(\rho_{t})$ Eq. (\ref{A5}) implies that  $\partial_{\tau} {\mathcal R}_{i} =0$: in this case the two terms 
at the right hand side simplify because $\partial_{i} \xi = (\partial \xi/\partial \rho_{t})\partial_{i} \rho_{t}$ and $\partial_{\tau}\xi 
= (\partial \xi/\partial \rho_{t})\partial_{\tau} \rho_{t}$.  Putting together the results obtained so far, we can therefore say that 
Eq. (\ref{A4}) is invariant for ${\mathcal R}_{i}(\tau, \vec{x}) \to {\mathcal R}_{i}(\tau, \vec{x}) + {\mathcal Q}(\vec{x})$ provided
$\xi$ is either a space-time constant or a function of the total energy density. 

The requirements of the previous paragraph correspond to the situation where Eq. (\ref{A2}) admit a fully 
inhomogeneous solution whose homogeneous limit is of quasi-de Sitter type. Let us therefore 
show explicitly that this is indeed the case in the simplest situation where $\xi$ is a space-time constant.
 Equations (\ref{A2}) implies the following decoupled equation\footnote{In Eq. (\ref{pos5}) we have chosen the geodesic slicing with $N=1$; in this case $\tau=t$ where $t$ denotes the 
cosmic time coordinate and the overdot denotes a derivation with respect to $t$.} for $K$:
\begin{equation}
\dot{K} - \frac{w+1}{2} K^2 = \frac{3}{2} \ell_{P}^2 \xi K,
\label{pos5}
\end{equation}
The general solution of Eq. (\ref{pos5}) is:
\begin{equation}
K(t,\vec{x}) = \frac{K_{0}(\vec{x}) \, e^{K_{\xi} [t - t_{0}(\vec{x})]/2}}{ (w+1) K_{0}(\vec{x}) \biggl[1 - e^{K_{\xi}[t - t_{0}(\vec{x})]/2}\biggr]  + K_{\xi} }, \qquad K_{\xi} = 3 \ell_{P}^2 \xi = \frac{1}{t_{\xi}},
\label{pos6}
\end{equation}
where $K_{0}(\vec{x})$ constant in time but not in space and $K_{\xi}$ is  just a parameter of the solution.  For $K_{\xi} [t - t_{0}(\vec{x})] \ll 1$ we have that $K(t,\vec{x})$ is singular while in the opposite 
limit it goes to a negative constant. In  the homogenous limit $t_{0}$ is constant also in space. 

Equations (\ref{pos5}) and (\ref{pos6}) show that when $\xi$ is a space-time constant we can derive an inhomogeneous 
solution whose homogeneous limit interpolates between a perfect fluid solutions and a quasi-de Sitter solution. 
These solutions are the inhomogeneous counterpart of various quasi-de Sitter solutions 
derivable in the fully homogenous limit \cite{BV}.  The generalization of Eq. (\ref{pos5}) to the case when 
$\xi= \xi(\rho_{t})$ is straightforward since the dependence on $\rho_{t}$ can be eliminated through Eq. (\ref{A2}). Some 
cases where $\xi$ has a power-law dependence on $\rho_{t}$ are even analytically solvable. 
We can therefore conclude that the large-scale curvature inhomogeneities are non-perturbatively 
conserved in numerous cases where the non-perturbative solution, in its fully homogeneous limit, admits a set of quasi-de Sitter backgrounds.

\subsection{Non-conservation of curvature perturbations}
Let us finally consider the case $\xi= \xi(\rho_{t},n_{t})$ where Eq. (\ref{A5}) becomes:
\begin{equation}
\partial_{\tau} {\mathcal R}_{i} = \frac{K}{3 (\rho_{t} + P_{eff})^2}  \biggl(\frac{\partial \xi}{\partial n_{t}}\biggr)\biggl[ 
\partial_{\tau} \rho_{t} \partial_{i} n_{t} - \partial_{i} \rho_{t} \partial_{\tau} n_{t}\biggr].
\label{A6}
\end{equation}
The term at the right hand side does not vanish, in general. To lowest order in the gradient expansion 
the diffusion current $\nu^{\alpha}$ does not contribute to the evolution of $n_{t}$. Equation (\ref{A6}) can then be rewritten by using Eqs. (\ref{EN}) and (\ref{dens}). The result is:
\begin{equation}
\partial_{\tau} {\mathcal R}_{i} = \frac{K^2 N}{3 (\rho_{t} + P_{eff})} n_{t} \biggl(\frac{\partial \xi}{\partial n_{t}}\biggr)\biggl[ \frac{\partial_{i} n_{t}}{n_{t}} - \frac{\partial_{i} \rho_{t}}{\rho_{t} + P_{eff}}\biggr] 
\label{A6a}
\end{equation}
Similar effects are expected when electromagnetic fields or scalar fields are present together with a dissipative fluid (see, for instance, 
the first two papers of Ref. \cite{sc1} for the case of scalar fields and the third paper of Ref. \cite{sc1} for the case of electromagnetic fields). The extension 
of the present considerations to a multicomponent viscous system is beyond the scope of this paper but it is conceptually feasible.

In summary, the viscous fluids do not necessarily jeopardize the large-scale conservation of the curvature inhomogeneities  
at least as long as $\xi$ is a function of the total 
energy density or of the trace of the extrinsic curvature. Conversely the large-scale conservation of the non-linear curvature perturbations is invalidated 
whenever  $\xi= \xi(\rho_{t}, n)$ [or when $\xi= \xi(K, n)$]. These observations have a counterpart in linear theory which will be discussed in the following section. 

\renewcommand{\theequation}{5.\arabic{equation}}
\setcounter{equation}{0}
\section{Back to linear theory}
\label{sec5}
\subsection{Scalar modes in linear theory}
In linear theory variables ${\mathcal R}_{i}$ and $\zeta_{i}$ defined in Eqs. (\ref{defR}) and (\ref{defZ}) have a well defined 
limit. Let us choose conformally Newtonian frame where the gauge freedom is removed and the coordinate 
system completely fixed
\begin{equation}
N^2(\tau,\vec{x}) = a^2(\tau) [ 1 + 2 \phi(\tau,\vec{x})],\qquad \gamma_{i j}(\tau,\vec{x}) = a^2(\tau)[ 1 - 2 \psi(\tau,\vec{x})]\delta_{ij},
\label{lim0}
\end{equation}
and expand Eqs. (\ref{defR}) and (\ref{defZ}) by assuming that $\phi$ and $\psi$ are both smaller than one.
The result of this limit is given by:
\begin{equation}
{\mathcal R}_{i} \to \partial_{i} {\mathcal R},\qquad \qquad \zeta_{i} \to   \partial_{i} \zeta,
\label{lim1}
\end{equation}
implying that ${\mathcal R}_{i}$ and $\zeta_{i}$ are, respectively, the spatial gradients of the curvature perturbation on comoving orthogonal 
hypersurfaces and of the density contrast on uniform curvature hypersurfaces, i.e.\footnote{The prime will denote a derivation with 
respect to $\tau$ and ${\mathcal H} =a^{\prime}/a$.}
\begin{equation}
{\mathcal R}= - \psi -  \frac{{\mathcal H} ( \psi^{\prime} + {\mathcal H} \phi)}{{\mathcal H}^2 - {\mathcal H}^{\prime}}, \qquad 
\zeta= - \psi + \frac{\delta\rho_{t}}{3 (\rho_{t} + P_{eff})}.
\label{lim1a}
\end{equation}
 In this section $\rho_{t}$ and $P_{eff}$ denote the background values of the corresponding quantity while $\delta\rho$ is the first-order fluctuation of the energy density and so on and so forth. In other words the conventions will be such that 
\begin{equation}
\rho(\tau, \vec{x}) = \rho_{t}(\tau) + \delta \rho_{t}(\tau,\vec{x}), \qquad \xi(\tau,\vec{x}) = \xi(\tau) +\delta\xi(\tau,\vec{x}).
\label{lim2}
\end{equation}
where $\delta$ will denote the first-order fluctuation of the corresponding quantity. The same conventions will be employed for all the other variables involved in the discussion. 
 
The linear order form of the evolution equation for ${\mathcal R}$ and $\zeta$ can be derived in perturbation theory and then compared with limit of Eq. (\ref{A4}). The result is is given 
by the following pair of equations where we have included, for the sake of comparison, also the terms that are of higher order 
in the gradient expansion but are consistent with the linearized approximation:
\begin{eqnarray}
{\mathcal R}^{\prime} &=& \frac{3 {\mathcal H} }{a (\rho_{t} + P_{eff})} \xi^{\prime}  ({\mathcal R} + \psi)- \frac{{\mathcal H}}{\rho_{t} + P_{eff}} \delta p_{nad} 
\nonumber\\
&+& \frac{3 {\mathcal H}^2}{a( \rho_{t} + P_{eff})} \delta \xi + \frac{\xi {\mathcal H} }{a(\rho_{t} + P_{eff})}\theta_{t} - \frac{3{\mathcal H} c_{st}^2 }{2\ell_{P}^2 (\rho + P_{eff})} \nabla^2 \psi,
\label{RPR}\\
\zeta^{\prime} &=& \frac{3 {\mathcal H} }{a(\rho_{t} + P_{eff})} \xi^{\prime} (\zeta + \psi)- \frac{{\mathcal H}}{\rho_{t}+ P_{eff}} \delta p_{nad} 
\nonumber\\
&+& \frac{3 {\mathcal H}^2}{a(\rho_{t} + P_{eff})} \delta \xi + \theta_{t} \biggl[\frac{{\mathcal H} \xi}{a(\rho_{t} + P_{eff})}  - \frac{1}{3}\biggr] 
\nonumber\\
&-& \frac{{\mathcal H}}{ 2 \ell_{P}^2 ( \rho_{t} + P_{eff})} \nabla^2 (\phi - \psi) - \frac{\xi}{a(\rho_{t} + P_{eff})}\nabla^2 \psi.
\label{zetapr}
\end{eqnarray}
In Eqs. (\ref{RPR}) and (\ref{zetapr}) the term $\theta_{t}$ denotes the three-divergence of the total velocity field.

Equations (\ref{RPR}) and (\ref{zetapr}) have been derived directly in the linear theory and they reproduce the results obtainable by linearizing . Consider then, for the sake of comparison, Eqs. (\ref{RPR}) and (\ref{A5}) in the limit 
$\delta p_{nad} \to 0$. From Eq. (\ref{RPR}) we will have 
\begin{equation}
{\mathcal R}^{\prime} = \frac{3 {\mathcal H} }{a (\rho_{t} + P_{eff})} \xi^{\prime}  ({\mathcal R} + \psi) +\frac{3 {\mathcal H}^2}{a(\rho_{t} + P_{eff})}\, \delta \xi,
\label{lim6}
\end{equation}
where $\theta_{t}$ has been neglected since it is of higher order in the gradients thanks to the momentum constraint (implying $\theta_{t} = \nabla^2({\mathcal R} + \psi)/{\mathcal H}$).

To see how things work in linear theory we can verify explicitly that the right hand side of Eq. (\ref{lim6}) vanishes  when $\xi= \xi(\rho_{t})$.
From the definition of $\zeta$ in linear theory we have that the momentum constraint can be expressed as:
\begin{equation}
\zeta = {\mathcal R} + \frac{\nabla^2\psi}{2 \ell_{P}^2 (\rho_{t} + P_{eff})}.
\label{lim7}
\end{equation}
Neglecting the gradients, Eq. (\ref{lim6}) is expressible,  in the case $\xi= \xi(\rho_{t})$, as:
\begin{equation}
{\mathcal R}^{\prime} = \frac{3 {\mathcal H}}{a(\rho_{t} + P_{eff})}\biggl(\frac{\partial \xi}{\partial\rho_{t}} \biggr) \biggl[ \rho_{t}^{\prime} ( \zeta + \psi) + {\mathcal H} \delta \rho_{t} \biggr]
\label{lim8}
\end{equation}
where we used that $\delta \xi = (\partial \xi/\partial \rho_{t}) \delta \rho_{t}$. But now thanks to the definition of $\zeta$ we have that $(\zeta+ \psi) = 3 (\rho_{t} + P_{eff}) \delta \rho_{t}$. Thus Eq. (\ref{lim8})  implies  ${\mathcal R}^{\prime} =0$ since, by covariant conservation of the background 
energy-momentum tensor, $\rho_{t}^{\prime} = - 3 {\mathcal H} ( \rho_{t}+ P_{eff})$.
This result does not hold if $\xi = \xi(\rho_{t},n_{t})$ so that, in general, the curvature perturbations induced by relativistic 
viscous fluids are not conserved. This derivation is the linear order counterpart of the discussion presented after Eq. (\ref{A5}).

\subsection{Standard gauge-invariant variables}
For infinitesimal coordinate transformations we have that 
$\phi\to \widetilde{\phi} = \phi - {\mathcal H} \epsilon_{0} - \epsilon_{0}'$ and  that $\psi \to \widetilde{\psi} = \psi + {\mathcal H} \epsilon_{0}$. The bulk viscosity coefficient transforms instead as:
\begin{equation}
\delta\xi \to \widetilde{\delta\xi} = \delta \xi - \xi^{\prime} \epsilon_{0}.
\end{equation}
Including the gradients in the appropriate 
entries of the perturbed metric (i.e. $\delta g_{ij} = 2 a^2( \psi \delta_{ij} - \partial_{i} \partial_{j} E)$ and $\delta g_{i0} = - a^2 \partial_{i} B$) 
and recalling that they transform as 
$B \to \widetilde{B} = B + \epsilon_{0} - \epsilon'$ and as $E \to \widetilde{E} = E - \epsilon$
the gauge-invariant fluctuations of the bulk viscosity fluctuations are:
\begin{equation}
\Xi = \delta \xi +\xi^{\prime} ( B - E^{\prime}), 
\end{equation}
while the Bardeen potentials are, as usual,  $\Phi = \phi + ( B - E')' + {\cal H} ( B - E')$ and  $\Psi= \psi - {\cal H} ( B - E')$.
In terms of the explicitly gauge-invariant fluctuations Eq. (\ref{RPR}) becomes 
\begin{eqnarray}
{\mathcal R}^{\prime} &=& \frac{3 {\mathcal H} }{a (\rho_{t} + P_{eff})} \xi^{\prime}  ({\mathcal R} + \Psi)- \frac{{\mathcal H}}{\rho_{t} + P_{eff}} \delta p_{nad} 
\nonumber\\
&+& \frac{3 {\mathcal H}^2}{a( \rho + P_{eff})} \Xi + \frac{\xi {\mathcal H} }{a(\rho_{t} + P_{eff})}\Theta_{t} - \frac{3{\mathcal H} c_{st}^2 }{2\ell_{P}^2 (\rho + P_{eff})} \nabla^2 \Psi,
\label{RPR2}
\end{eqnarray}
where $\Theta_{t} = \theta_{t} + \nabla^2 E^{\prime}$ is the gauge-invariant expression of the total velocity field. 
\subsection{Fluctuations of the expansion and $\delta {\mathcal N}$ formalism}
We can also compute the total expansion rate with the aim of showing that the presence of the bulk viscosity affects  the basis 
of the so-called $\delta {\mathcal N}$ formalism stipulating that $\zeta$ can be related to the scalar-field perturbations at the initial time computed in the 
uniform curvature gauge once we know the derivatives of the number of efolds with respect to the initial values of the unperturbed scalar fields and their derivatives. 

Let us therefore introduce the inhomogeneous generalization of the total number of efolds 
\begin{equation}
{\mathcal N}(\vec{x},\tau_{*}, \tau_{f}) = \frac{1}{3} \int_{\tau_{*}}^{\tau_{f}} \nabla_{\alpha} u^{\alpha} \, N \, d\tau,
\label{dN1}
\end{equation}
where $\tau_{*}$ denotes the initial time;all the quantities of the integrand are space-time dependent. We recall that in linear 
theory we can define 
\begin{equation}
\nabla_{\mu} u^{\mu} = (\nabla_{\mu} u^{\mu})^{(0)} + \delta^{(1)}(\nabla_{\mu} u^{\mu}) + \delta^{(2)}(\nabla_{\mu} u^{\mu}) +\,...
\label{dN1a}
\end{equation}
Without committing ourselves to a specific gauge choice, from Eq. (\ref{dN1a}) and from the fluctuations of the lapse function, Eq. (\ref{dN1}) becomes:
\begin{eqnarray}
{\mathcal N}(\vec{x},\tau_{*}, \tau_{f}) &=& \overline{{\mathcal N}}(\tau_{*},\tau_{f}) + \frac{1}{3} \int_{\tau_{*}}^{\tau_{f}} (\theta_{t} + \nabla^2 E') \, d\tau
- \int_{\tau_{*}}^{\tau_{f}} \psi^{\prime} \, d\tau,
\nonumber\\
\overline{{\mathcal N}}(\tau_{*},\tau_{f}) &=& \int_{\tau_{*}}^{\tau_{f}} \,{\mathcal H} d \tau =  \int_{t_{*}}^{t_{f}} H d t,
\label{dN2}
\end{eqnarray}
where ${\mathcal H} = H a$ and $dt = a d\tau$.
From the first-order fluctuation of the covariant conservation of the total energy-momentum tensor we can express $\psi^{\prime}$ as
\begin{eqnarray}
\psi^{\prime} &=& \frac{1}{3} [ \theta_{t} + \nabla^2 E^{\prime}] + \frac{\delta \rho_{t}^{\prime} + 3 {\mathcal H} ( \delta \rho_{t} + \delta p_{t})}{3 ( \rho_{t} + P_{eff})} 
- \frac{3 {\mathcal H}^2 }{a ( \rho_{t} + P_{eff})} \delta\xi
\nonumber\\
&-& \frac{{\mathcal H} \xi}{a(\rho + P_{eff})}[ \theta_{t} + \nabla^2 E^{\prime} - 3 ( \psi^{\prime} + {\mathcal H} \phi)].
\label{dN3}
\end{eqnarray}
Inserting Eq. (\ref{dN3}) into Eq. (\ref{dN2})  we can easily obtain 
\begin{eqnarray}
{\mathcal N}(\vec{x},\tau_{*}, \tau_{f}) &=& \overline{{\mathcal N}}(\tau_{*},\tau_{f}) +  \int_{\tau_{*}}^{\tau_{f}} \biggl\{ \frac{3 {\mathcal H}}{a (\rho_{t} + P_{eff})} [ {\mathcal H} \delta \xi + \xi^{\prime} ( \psi + \zeta)] - (\zeta^{\prime} + \psi^{\prime}) 
\nonumber\\
&+& \frac{\xi}{a (\rho_{t} + P_{eff})} \nabla^2 {\mathcal R} - \frac{{\mathcal H}}{(\rho_{t} + P_{eff})} \delta p_{nad}\biggr\},
\label{dN4}
\end{eqnarray}
where, according to Eq. (\ref{lim1a}), we used that $\delta \rho = 3 (\zeta + \psi) (\rho + P_{eff})$.
When  $\delta p_{nad}\to 0$ and $\xi= \xi(\rho_{t})$ we have:
\begin{equation}
{\mathcal N}(\vec{x},\tau_{*}, \tau_{f}) = \overline{{\mathcal N}}(\tau_{*},\tau_{f}) -  \int_{\tau_{*}}^{\tau_{f}} \biggl( \zeta^{\prime} + \psi^{\prime} \biggr) d\tau,
\label{dN5}
\end{equation}
implying 
\begin{equation}
{\mathcal N}(\vec{x},\tau_{*}, \tau_{f}) = \overline{{\mathcal N}}(\tau_{*},\tau_{f})+   [\zeta(\vec{x},\tau_{*})+ \psi(\vec{x},\tau_{*})]  -  [\zeta(\vec{x},\tau_{f})+ \psi(\vec{x},\tau_{f})] .
\label{dN6}
\end{equation}
Let us then evaluate $(\zeta + \psi)_{\tau_{f}}$ in the uniform density gauge (i.e. $\delta \rho_{t} =0$) while $(\zeta + \psi)_{\tau_{*}}$ is evaluated 
in the uniform curvature gauge (i.e. $\psi =0$).  This choice implies that:
\begin{equation}
 [\zeta(\vec{x},\tau_{f})+ \psi(\vec{x},\tau_{f}) ]\to 0, \qquad
  [\zeta(\vec{x},\tau_{*})+ \psi(\vec{x},\tau_{*})] = \frac{\delta\rho_{t}}{3 ( \rho_{t} + P_{eff})}.
\label{dN7}
\end{equation}
Thus, from Eqs. (\ref{dN6}) and (\ref{dN7}) we have that 
\begin{equation} 
\delta {\mathcal N} =   {\mathcal N}(\vec{x},\tau_{*}, \tau_{f}) - \overline{{\mathcal N}}(\tau_{*},\tau_{f})= \frac{\delta\rho_{t}}{3 ( \rho_{t} + P_{eff})}.
\label{dN8}
\end{equation}
The same formulas derived above hold in the single scalar field case (at least up to some point) by setting $\xi =0$ and by recalling 
that, in the single scalar field case, ${\mathcal R} = - \psi - ({\mathcal H}/\varphi^{\prime}) \delta \varphi$ where 
$\delta\varphi$ denotes the scalar field fluctuation. Thus, in the single scalar field case, we will have 
\begin{equation}
[\zeta(\vec{x},\tau_{*})+ \psi(\vec{x},\tau_{*})] =\biggl[ {\mathcal R}_{*}  + \psi_{*} + \frac{\nabla^2 \psi_{*}}{ 2 \ell_{P}^2 (\rho_{\varphi} + p_{\varphi}) }\biggr] \to - \frac{{\mathcal H}}{\varphi^{\prime}} \delta \varphi.
\label{dN9}
\end{equation}
In analogy with Eq. (\ref{dN8}), Eq. (\ref{dN9}) can be expressed as $\delta {\mathcal N} \simeq - (\partial {\mathcal N}/\partial \varphi)_{*} \delta \varphi(\vec{x},\tau_{*})$, which is the standard result of the single scalar field case. 

We can therefore conclude that the presence of bulk viscous stresses affects the $\delta {\mathcal N}$ only if $\xi = \xi(\rho_{t},n_{t})$. Conversely, 
if $\xi = \xi(\rho_{t})$ both ${\mathcal R}$ and $\zeta$ are constant and $\delta {\mathcal N}$ only depends on the 
values of $\zeta + \psi$ between the initial and final times characterizing the integrated (total) expansion.
\renewcommand{\theequation}{6.\arabic{equation}}
\setcounter{equation}{0}
\section{Concluding remarks}
\label{sec6}
The hypothesis of reversibility of the plasma has been dropped by allowing for a relativistic dissipative fluid as the dominant source of curvature inhomogeneity in the non-perturbative regime.  In the general situation the non-linear evolution does not  preserve the curvature inhomogeneities.  However if the bulk viscosity is either a space-time constant or if it depends solely on the 
total energy density of the plasma, then a further symmetry 
prevents the growth of curvature inhomogeneities. To lowest order in the gradient expansion the dependence on the energy density 
can be traded for a dependence on the trace of the extrinsic curvature. In all these cases the curvature inhomogeneities
are non-perturbatively conserved. 
The validity of the linear theory has not been posited as a necessary requirement of the derivation. Nonetheless the non-perturbative 
results have a perturbative counterpart describable in terms of the conventional gauge-invariant variables 
namely the curvature perturbations on comoving orthogonal hypersurfaces and the 
density contrast on uniform curvature hypersurfaces. 

The only contribution to the non-perturbative evolution of curvature inhomogeneities comes from the 
bulk viscous stresses while the shear viscosity affects the evolution of the 
traceless part of the extrinsic curvature and it appears to higher order in the gradient expansion. 
The only contribution to the inhomogeneous deceleration 
parameter comes from the bulk viscosity. Beyond linear theory the 
large-scale acceleration can only be driven by the bulk pressure  
and no acceleration can take place thanks to the shear viscosity. 

The results of this investigation confirm that the non-linear conservation of curvature perturbations is 
non a generic phenomenon. However when the curvature inhomogeneities are conserved
the evolution equations of the extrinsic curvature can be solved in fully inhomogeneous terms. These 
solutions correspond, in the homogeneous and isotropic limit, to a quasi-de Sitter stage of expansion. 
One could therefore speculate that the non-perturbative constancy of the curvature inhomogeneities 
pins  down  a class of viscous coefficients and, ultimately, a specific set of physical properties of the geometry. 
It is amusing that this logic is opposite to the one commonly pursued in linear theory where a quasi-de Sitter stage 
of expansion is postulated to insure, somehow, the perturbative conservation of the linearized fluctuations.

\end{document}